# Quantum Networks for Open Science Workshop

**Office of Advanced Scientific Computing Research
Department of Energy**

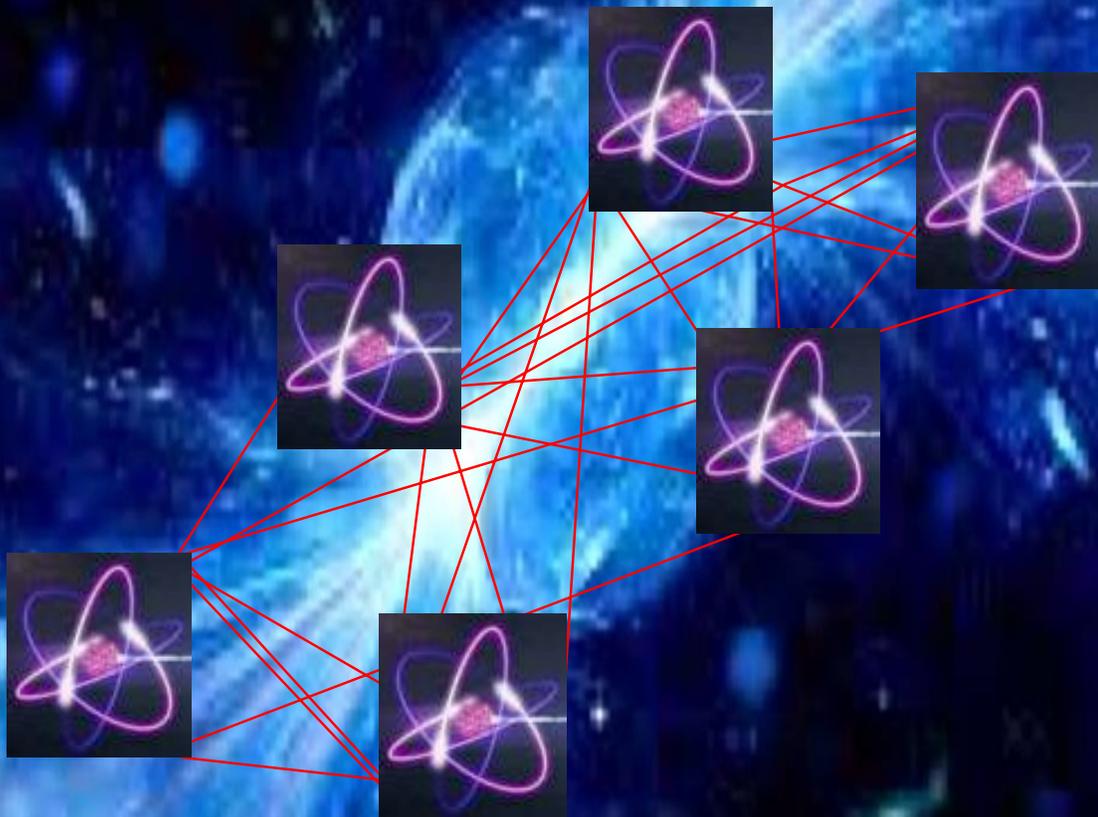

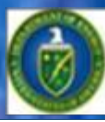
U.S. DEPARTMENT OF ENERGY | Office of Science

**Program Manager:**
**Thomas Ndousse-Fetter (DOE/ASCR)**

**Workshop POC:**
**Nicholas Peters (Oak Ridge National Laboratory)**

**Workshop Committee:**
**Warren Grice (Qubitekk, Inc.)**
**Prem Kumar (Northwestern U.)**
**Tom Chapuran (Perspecta Labs)**
**Saikat Guha (University of Arizona)**
**Scott Hamilton (MIT LL)**
**Inder Monga (LBL)**
**Ray Newell (LANL)**
**Andrei Nomerotski (BNL)**
**Don Towsley (UMass)**
**Ben Yoo (UC Davis)**

**Workshop Attendees**
**(See Appendix - A)**

Table of Contents

**List of Abbreviations**
- ASCR – Advanced Scientific Computing Research
- ATM – Asynchronous Transfer Mode
- BES – Basic Energy Sciences
- CPU – Central Processing Unit
- CWDM – Coarse Wavelength Division Multiplexing
- CV – Continuous Variable
- CV-QKD - Continuous Variable Quantum Key Distribution
- DOE – Department of Energy
- DWDM – Dense Wavelength Division Multiplexing
- DV – Discrete Variable
- EDFA – Erbium Doped Fiber Amplifier
- E-O – Electro to Optical
- FES – Fusion Energy Sciences
- GMCS – Gaussian Modulated Coherent State
- GMPLS –Generalized Multiprotocol Label Switching
- GPU – Graphics Processing Unit
- HEP – High Energy Physics
- HPC – High Performance Computing
- IP – Internet Protocol
- LAN – Local Area Network
- MAN – Metropolitan Area Network
- MPLS – Multi-Protocol Lambda Switching
- Mux/DeMux – Multiplexer/De-Multiplexer
- NASA – National Aeronautics and Space Administration
- NIST – National Institute of Standards and Technology
- O-E – Optical to Electrical
- O-E-O – Optical to Electrical to Optical
- OSI – Open Systems Interconnection
- QoE – Quality of Entanglement
- QBER – Quantum Bit Error Rate
- Q-HPCC – Quantum High-Performance Computing and Communications
- QIS – Quantum Information Science
- Q-Internet – Quantum Internet
- QKD – Quantum Key Distribution
- Q-LAN – Quantum Local Area Network
- Q-MAN – Quantum Metropolitan Area Network
- QNOS – Quantum Networks for Open Science
- Q-ROADM – Quantum Reconfigurable Optical Add-Drop Multiplexer
- Q-SAN – Quantum Storage Area Network
- Q-WAN – Quantum Wide Area Network
- ROADM – Reconfigurable Optical Add-Drop Multiplexer
- SS7 – Signaling System 7
- SQL – Standard Quantum Limit
- SAN – Storage Area Network
- SONET – Synchronous Optical Networking
- TCP – Transmission Control Protocol
- WDM – Wavelength Division Multiplexing
- WAN – Wide Area Network

# Executive Summary

Quantum computing systems currently being developed will have extraordinary capabilities to effectively solve complex problems in computational sciences, communication networks, artificial intelligence, and data processing, and will provide a powerful capability for researchers in almost every scientific discipline. Harnessing the full potential of quantum computing will require an ecosystem with a broad spectrum of quantum technologies. Quantum networks are one of the critical and highly anticipated components of this ecosystem. The combination of quantum computing and quantum networks are crucial to the US Department of Energy's (DOE) mission to provide scientists with the state-of-the-art computational capabilities.

DOE leadership has led, not only to some of the most powerful high-performance computing systems, but also to state-of-the-art high-performance networks that have brought major contributions to modern internet technologies. The fact that DOE innovation in communications networks has paralleled the growth of high-performance computing (HPC) is not a coincidence. Digital computing and communications/networking have evolved in parallel and have leveraged one another since the inception of the modern computing ecosystem. Innovations in communications/networking technologies have led to the design, deployment, and operation of advanced supercomputers. Given that the DOE science environment consists of geographically distributed computing resources, science facilities, and research teams, it is highly likely that the deployment of quantum systems will be similarly distributed. It follows that quantum networks will be critical to access and share these distributed quantum systems and it is anticipated that a similar coevolution strategy will be adopted in setting the strategic funding and research priorities for quantum computing and quantum communications/networking.

DOE envisions a quantum networking ecosystem that will embody the capabilities needed to support a highly diversified QIS portfolio, namely scalable and adaptable quantum network infrastructures designed to support the transmission of diverse types of quantum information (discrete, continuous, or hybrid quantum states). It is anticipated that new quantum networks will be designed to coexist with its existing Energy science network (ESnet), a high-performance optical backbone network connecting DOE's scientific resources. Although there are many parallels between the quantum and classical versions of networks and computing, the unique character of quantum information presents some formidable challenges for the development of a quantum network. Quantum information, which is encoded in quantum objects, cannot be amplified or duplicated; and quantum states are altered if measurements are performed on them. Thus, common tasks on the classical internet such as routing and buffering will have to be performed in a completely different way on the quantum internet. In addition, quantum networks will be limited in scale until a viable quantum repeater technology becomes available. Nevertheless, the decades of innovation that have led to today's internet should guide the development of the quantum internet.

The long-term benefits of quantum networks go far beyond interconnecting distributed quantum information resources and opening new avenues for secure communications, yielding a paradigm shift in the concept of modern telecommunications and, eventually, will lead to a quantum internet interacting and coexisting with the current classical internet.  DOE's approach to quantum networks for open science should reflect this long-term vision of developing quantum high-performance computing and communications that will not only support its science mission but also contribute to the emerging quantum internet.

# 1 Quantum Networks for Open Science (QNOS)

## 1.1 Introduction

We now live in a networked society powered by computing and connected by a vast communications network, commonly known as the internet. This partnership of computing and communications networks enables economic prosperity and public wellbeing by facilitating business, collaborations, sharing, and access to healthcare. Indeed, nearly every aspect of modern life is impacted by both computing and networks, and it is difficult to imagine either one without the other. Quantum information technologies, in which the laws of quantum mechanics govern the control and transmission of information, promise revolutionary new capabilities that are fundamentally different from what will be possible with advances in classical technology alone. Quantum networks are a critical and enabling resource to bring quantum technologies and their benefits to society. More specifically, harnessing quantum technology to support the emerging quantum computing ecosystem in the science community is a subject of intense investigation in major academic and national laboratories worldwide. DOE has traditionally provided leadership in high-performance computing and communications to support open science. Quantum networks will interconnect quantum information held in various types of quantum computing systems by converting such information into a photonic form suitable for transmission over optical communications infrastructures that preserve the quantum information while in transit.

It is in this context that DOE convened the Quantum Networks for Open Science (QNOS) Workshop in September 2018. The workshop was primarily focused on quantum networks optimized for scientific applications with the expectation that the resulting quantum networks could be extended to lay the groundwork for a generalized network that will evolve into a quantum internet (Q-Internet). The QNOS Workshop complements and extends a series of workshops and roundtables organized by DOE in recent years (ASCR [1], ASCR [2], HEP [3], BES [4], and FES [5]), developing a Quantum Information Sciences (QIS) portfolio across its major science programs. There is a consensus in the community that there are quantum applications in DOE's QIS portfolio, which will require quantum networks. The primary objective of the QNOS workshop, which included a diverse set of participants from the quantum physics, telecommunications engineering, optical communications, and computer science communities, was to explore the challenges and opportunities associated with developing quantum networks to enable these distributed quantum applications.

## 1.2 Workshop Scope

Quantum networks are a disruptive technology with competing implementation approaches. There are several quantum computing approaches (such as those based on superconducting qubits, trapped-ion qubits, or other emerging qubit technologies) to host the underlying quantum bits or qubits which are used to harness the fundamental power of quantum mechanics. Qubits are analogous to binary bits in classical digital computing but, whereas a classical bit is described by a single on or off value, a qubit is described by a pair of complex numbers. These properties give quantum computers extraordinary capabilities to tackle complex scientific problems that are thought to be intractable with classical supercomputers. While different quantum system types may be better suited for different QIS applications, only photonic systems are suitable for transmitting quantum information over long distances. Photons provide a natural source of 'flying' quantum information carriers with long coherence times, for communications across free space or over guided optical fiber links.

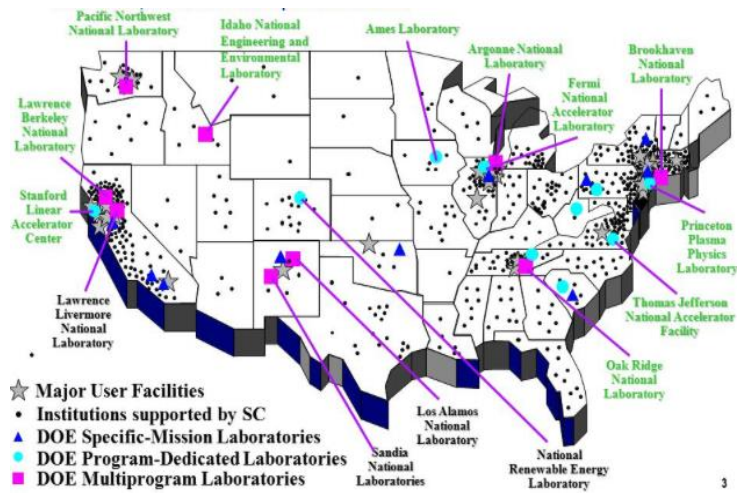

**Figure 1**: DOE Distributed Science Environment

Additionally, mature transparent optical communications technology developed for commercial telecommunications systems such as Dense Wavelength Division Multiplexing (DWDM), Reconfigurable Optical Add/Drop Multiplexers (ROADM), FlexGrid [6], forward error correction, Generalized Multiprotocol Label Switching (GMPLS) signaling protocols, and wavelength routing algorithms could be leveraged to develop transparent optical quantum networks.

The vision and scope of quantum networking for the QNOS workshop, as articulated in the charge letter to the organizing committee, was to identify the opportunities and challenges for developing scalable quantum networks by transmission of quantum information through optical fiber. This quantum networking approach is referred to as optical quantum networking in the remainder of this document to define the scope of the present document to exclude approaches based on free-space (for example, based upon transmission to and from satellites), or other transmission media. These constraints and other DOE mission priorities and investments helped shape the scope, context, and guidelines for the QNOS workshop as follows.

- The development of optical quantum networks should leverage DOE's optical fiber plants used by its flagship high-performance optical backbone Energy Sciences Network (ESnet) [7].
- The development of open network architectures and protocols should support multi-vendor components adaptable to new quantum information science applications without major redesign.

- The resulting networking approach should be scalable and be capable of accommodating new network domains as needed.
- The focus of this workshop should be limited to quantum communications networks over an optical fiber transmission medium. While future fiber and free-space quantum networks are expected to converge, the latter is out-of-scope for the purposes of this report.

Within this scope, the workshop was organized to facilitate discussions aimed at highlighting the key challenges in the development of a quantum network to support open science applications. The discussions focused on the technological advances that would be needed to cover a wide range of topics including new science capabilities that would be enabled by quantum network technology, quantum network architecture, and quantum network devices and subsystems, and quantum network operations and management strategies.

## 1.3 Motivating Drivers and Science Applications for Quantum Networks

DOE has traditionally pushed the limits of information technologies to support cutting-edge science. This tradition will likely continue with quantum information technology. This means that DOE's quantum networks and, to a large extent, its quantum computing systems, will be uniquely optimized for high-performance scientific applications. Quantum Information Science (QIS) is poised to usher in a new science frontier and DOE, with its highly distributed science environment as shown in Figure 1, is set to explore new opportunities with QIS initiatives across its major science programs. DOE's main motivation in quantum networks is to enable the efficient use of distributed quantum resources and to enable a new generation of scientific applications that can harness quantum mechanical properties such as entanglement, squeezing, superposition, and teleportation to accelerate scientific breakthroughs in core science missions. Many aspects of quantum information science such as quantum sensing, distributed quantum simulation, and many-body quantum entanglement, have direct bearing on DOE interests including, for example, materials modeling, molecular dynamics, and high-energy physics.

### 1.3.1 Quantum Sensing

Quantum technologies manipulate individual quantum states and make use of superposition, entanglement, squeezing, and backaction evasion. Quantum sensors [3, 8] exploit these phenomena to make measurements with a precision better than the Standard Quantum Limit (SQL), with the ultimate goal of reaching the Heisenberg Limit. A single quantum sensor can only take advantage of quantum correlations in a single location, while a quantum network could exploit the correlations across an array of sensors, linking them to each other with quantum mechanical means. This improves the sensitivity and scalability of the resulting entangled system simultaneously allowing it to benefit from the long-distance baseline between the sensors. Below we list several options for possible quantum networks with quantum sensors and give examples of how these systems can be used for science applications:

- **Quantum network of atomic clocks**: A standalone multi-atom atomic clock already demonstrates stability close to the SQL, which is set by the number of atoms and integration time. In a quantum network, multiple atomic clocks can be connected together providing superior clock synchronization and stability [9, 10]. Compared to a single clock, the ultimate precision will improve as 1/K, where K is the number of clocks [9]. If the same clocks are connected via a classical network, the precision scales as $1/\sqrt{K}$. Thus, connecting 100 clocks with a quantum network potentially improves precision by a factor of 10 compared to connection by classical networking.

- **Phase-sensitive quantum network**: Quantum networks could transfer the sensor phase information between different locations. An example of this is interferometry over long distances enabled with quantum repeaters [11]. The phase difference is measured by transmitting the quantum states between remote locations using quantum teleportation. The concept was discussed in the context of long-baseline optical telescopes to improve the angular resolution [12]. However, this approach could be generalized to improve sensitivity for different observables.

- **Quantum network of magnetometers**: A network of entangled magnetometers based on atomic systems could have very high sensitivity to external magnetic fields due to large quantities of coherent atoms and long-distance baselines [13].

The above improvements in quantum sensing are enabled by quantum networks and will allow for better sensitivity, for example, in Dark Matter searches, by looking for transient changes in fundamental constants leading to desynchronization of the clock nodes [14]. The long-distance baselines allowed by the quantum networks could dramatically improve sensitivity for axion searches and other measurements probing variations of light polarization [15]. Coherent interactions of dark sector particles could be measured employing a quantum network of quantum memories. Here a new dark-sector particle interacting coherently with the spin of the atomic memories (e.g., [16]) may alter the state of the dark-state polariton wave front. Last but not least, optical telescopes connected across the globe in a quantum network could allow determination of apparent positions of stars with unprecedented precision [17]. Evolution of the above ideas will depend very much on theoretical efforts to develop concepts of experiments and to evaluate their sensitivity. As these techniques mature over time, they should be included in the DOE planning process for new experiments.

### 1.3.2 Distributed Quantum Computing

Computing models, in which several computing systems interconnect by Storage Area Networks (SANs), Local Area Networks (LANs), or Wide Area Networks (WANs), to perform global computations, are quite common in in classical digital computing. One of the main motivations for interconnecting small quantum computers through the types of quantum networks discussed in this report is the exponential computational speed-up. This is significant given the daunting challenge to develop large-scale quantum systems (current state-of-the-art is ~ 70 qubits). For these reasons, it has been proposed [18, 19] that quantum supercomputing can be achieved by using quantum networks (Quantum Storage Area Networks (Q-SANs), Quantum Local Area Networks (Q-LANs), or Quantum Wide Area Networks (Q-WANs)) to interconnect quantum systems containing a modest number of qubits. For example, consider two isolated quantum systems with 3 qubits each. The quantum state of each system, expressed as a density matrix, is a proxy for how complicated a system can be simulated, and is described by $2^{2*3}-1 = 63$ independent real parameters. Doubling the number of systems by classical networking results in a product state of the two individual systems, approximately doubles the number of free parameters needed to describe the joint system. However, by interconnecting the systems with a dense quantum network, the combined system is described by $2^{2*2*3}-1 = 4095$ real numbers, thus illustrating the exponential advantage of quantum networking. Different quantum computing applications require varying quantum resources and potentially different types of quantum operations for implementation. If a large enough number of quantum computational resources are not available in a single physical platform, or if different platforms excel at different operations (in analog to using both Central Processing Units (CPUs) and Graphics Processing Units (GPUs) in HPC) then distributed quantum computing may enable dramatic improvements in computational capability over classically networked quantum computational resources. Distributed quantum computing also has an advantage of improved resilience. Should one quantum computing node be unavailable, other nodes can be networked to replace it.

### 1.3.3 Blind Quantum Cloud Computing

Cutting edge quantum computing resources will invariably be shared among many users. This raises the question of how one can secure a particular computing task from, for example, a malicious user. Even if all users can be trusted, the integrity and privacy of what is being computed may need to be protected from users with joint access to quantum computing resources. This type of computing model is possible through blind quantum computing [20, 21], whereby users can outsource quantum computing tasks to a quantum server, possibly in a remote location. This can be done in a way that maintains the secrecy of both the computational commands and the results. The computations are kept secret not only from outside observers, but also from the quantum computer that carries out the computation. The concept of blind quantum computing was originally devised for cases in which quantum computing resources may be untrusted. However, the method has obvious applicability to a deployment model in which computing resources that are concentrated in a relatively small number of locations are meant to serve users in many different locations. The advantages are particularly appealing for computations that include sensitive data. Such methods could be useful for example, in biological computing on protected health data.

### 1.3.4 Quantum Key Distribution

Quantum Key Distribution (QKD) is one of the earliest and most advanced fields of quantum information science. QKD networks [22] however, are more limited than the vision for generic quantum networks discussed in this report. QKD transmits quantum photonic states between two users such that, after appropriate post-processing, a shared secret key can be established between the two users. In typical use cases, most transmitted photons are not received due to transmission loss. QKD may be performed with or without the use of quantum entanglement. Its security is rooted in the physical laws of quantum states and their measurement. Unlike classical cryptographic techniques, advances in mathematical algorithms and in computing technologies do not impact the security of QKD. For point-to-point QKD through a bosonic channel (such as optical fibers), there is a fundamental limit of achievable secure key rate for a given transmission loss value. To get around this rate-loss trade off, one can either employ a trusted relay node, Twin-Field QKD, or a quantum repeater (covered later in this report). The state-of-the-art distances for fiber-based QKD are achieved with the trusted node. Each trusted node typically contains two halves of a QKD system, the first exchanges keys to one neighboring node, and the second exchanges keys with a different neighboring node. The keys from neighboring directions are combined classically at the center node, for example, with a logical "XOR" function, and the combination is sent out to the neighboring nodes. By using the secret keys shared with the central node, the two end nodes now can decrypt the key sent by the central node establishing a shared key. The primary disadvantage of the trusted node is that it does not allow more general quantum communications protocols, such as end-to-end quantum data transmission. Protocols related to QKD exist which can distribute quantum digital signatures from one entity to multiple entities to later authenticate messages.

### 1.4 Challenges and Opportunities

**Challenges**

- Developing distributed computational science applications on new platforms is time-consuming, costly, and high risk for domain scientists, especially for a disruptive platform such as quantum networks, which are not mature.
- Lack of quantum computing platforms ready to be networked to enable distributed computing.
- Quantum networking resources such as entanglement and teleportation are still not well understood among domain scientists who could exploit them to solve new classes of scientific problems.
- Any attempt to exploit the capabilities of quantum networks in the development of science applications is limited to simulation, laboratory experiments, or to a network of modest distance scales because there are currently no repeater-based quantum networks.

**Opportunities (Short-Term)**

- Identify the quantum network functionality that is needed to enable new science applications (long baseline telescopes, Heisenberg-limited interferometry, improved clock synchronization,), e.g., qubit, qudit, vs. continuous variables, what degrees of freedom, two-qubit vs. multi-qubit entanglement, etc.
- Identify the testable parameter regimes accessible as quantum sensor networks improve scale and precision.

**Opportunities (Long-Term)**

- Make quantum sensor networks available to test new theories/probe for new physics.
- How does one use protocols like blind quantum computing to provide privacy of sensitive data yet ensure quantum computation is only for authorized applications?

# 2 Towards Quantum Networks in Open Science Environment

Networks have become indispensable in modern digital society. The DOE operates a network called ESnet [7], shown in Figure 2, to interconnect critical scientific resources and enable distributed research teams to share information. ESnet is a high-performance backbone network with a capacity around 20 petabytes of data monthly with some 13,000 miles of coast-to-coast dedicated fiber. It supports robust network services and disruptive research and development to deliver capabilities that are not commercially available. The information transmitted over ESnet and all similar networks including the internet is classical in the sense that it is encoded as a string of bits, each of which is in one of the two definite states, either 0 or 1. However, in the emerging quantum networks the information transmitted can be represented by quantum superposition states, such as a qubit, given by $\alpha|0\rangle + \beta|1\rangle$, where $\alpha$ and $\beta$ are complex numbers, usually normalized so the total measurement probability is one. While classical networks are designed to carry digital information that can be stored, processed, and re-transmitted while in transit, quantum networks will carry quantum photonic states that are subject to strict quantum mechanical constraints. Among the constraints, illustrated specifically for qubits, are the following: a) qubits are fragile in the sense that they can decohere or be lost when they interact with their transmission medium, therefore they are relatively short-lived, and managing them in wide-area scale distances poses serious challenges; b) no-cloning: quantum mechanics imposes a strict restriction on copying or cloning of qubits – an operation that is very common in classical networking; and c) of the many ways a qubit may be realized, not all are suitable for transmission over large distances. These constraints and other fundamental network challenges discussed in this report have not been adequately addressed by the research community.  To date, in comparison with classical networks, relatively limited progress has been made on quantum networks. By far, the most common configurations use simple point-to-point links, with one of two users "Alice and Bob" on each end of a fiber link.  This configuration is limited to links of at most a few hundred kilometers without a quantum repeater. The use of these simple networks has been primarily to demonstrate QKD networking concepts. The focus of this workshop was on the opportunities and challenges to be addressed in the research and development of multi-user wide-area scale quantum networks, such as in Figure 1, that have the potential to become the Q-Internet.

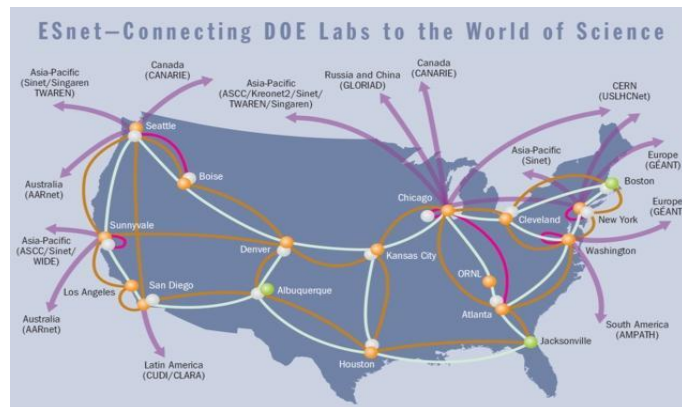

**Figure 2**: DOE High-Performance Optical Backbone Network to be leveraged for DOE's Q-WAN

## 2.1 Photonic Quantum Networks

The development of transparent optical networks has emerged as a critical enabler for quantum networking. Transparent light paths are essential for end-to-end transmission of photons carrying quantum states, as each optical to electrical (O-E) or electrical to optical (E-O) conversion along a non-transparent light path would also correspond to a quantum-to-classical (classical-to-quantum) conversion. The ultimate goal is a quantum network that distributes quantum states among various nodes in the network, while retaining their quantum properties with high fidelity. There are two basic approaches to achieve this, send quantum states directly, or pre-distribute entanglement, which is then used for quantum teleportation of a quantum state. Additionally, although entanglement has been demonstrated for a variety of qubit types, there are very few options for the

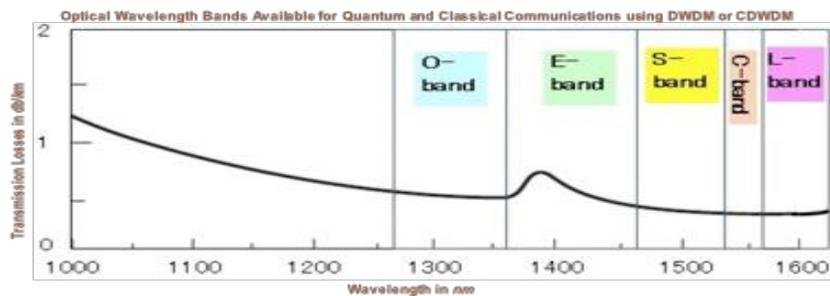

**Figure 3**: Optical Fiber Low Loss Transmission Spectrum

generation and distribution of entanglement over a wide area to support geographically distributed quantum systems. Entanglement carried by optical photons (photonic entanglement) appears to be one of the most promising options for long-distance fiber transmission due to its abundant bandwidth, low-noise properties [23]. In the near future, entangled sources should be designed such that they are easily integrated with deployed telecommunications optical fiber systems. Its low transmission losses, low cost, flexibility, and abundant capacity are made possible with multiplexing schemes such as DWDM and Coarse Wavelength Division

Multiplexing (CWDM). The wavelength region from 1260 nm to 1625 nm, which can be divided into five wavelength bands referred to as the O, E, S, C and L bands, as shown in Figure 3, has low transmission loss and is commonly utilized for telecommunication transmission. Optical quantum network designers would likely benefit from the emerging optical spectrum management scheme called FlexGrid [6] which breaks the spectrum up into small (typically 12.5 GHz) slots, but dynamically assigns contiguous slots that can be joined together to form arbitrary sized blocks of spectrum on demand.

Despite these attractive features of optical fiber systems for modern telecommunications, their ability to carry entangled pairs or quantum states over long-distances remains largely underdeveloped. Fiber and optical component loss represent significant challenges reducing transmitted rates. Optical fibers used in modern telecommunications are designed and optimized for long-distance transmission and maximum information carrying capacity. In order to achieve these objectives, designers must deal with linear (cross-talk, polarization) impairments and non-linear (four-wave mixing, stimulated Raman and stimulated Brillouin scattering) impairments that impact signal propagation through fiber. Using these fiber systems to carry quantum states, which have different requirements, introduces additional design challenges, especially if classical signals and quantum states coexist in the same fiber systems. A conventional Wavelength Division Multiplexing (WDM) switched optical telecommunication network link re-engineered for coexistence of quantum information and classical network traffic is shown in Figure 4. A quantum repeater is used to extend the reach of the link while an Erbium-Doped Fiber Amplifier (EDFA) fulfils a similar function for the classical optical signals.

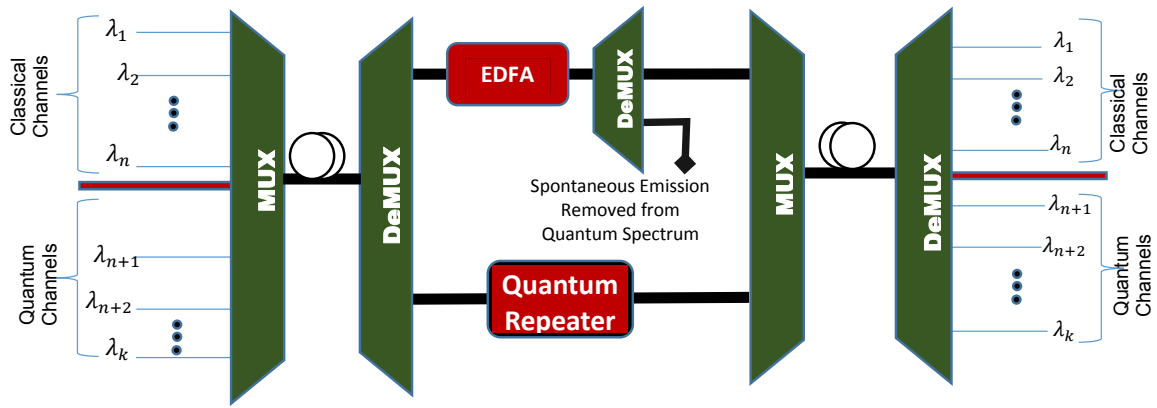

**Figure 4**: Optical communications fiber link engineered with two wavelength partitions to simultaneously carry photonic quantum states and classical traffic.

## 2.2 Quantum Networks over Telecommunications Optical Fiber Systems

Today's optical communications rely on Optical-Electrical-Optical (O-E-O) conversion for end-to-end management and control. In this mode, the entire optical payload carried on each wavelength is converted to an electronic digital signal for processing (such as routing, error correction, signal regeneration, and flow control) and then converted back to optical for transmission. This is referred to as "Opaque Optical Networking." This type of optical networking presents a challenge if the optical payloads are quantum states as they cannot be faithfully regenerated after a measurement. An approach more suitable for quantum networking is transparent optical networking in which the end-to-end signal remains in the optical domain [24, 25]. In transparent or all-optical optical networking, only optical devices are employed along the entire end-to-end light path and the channel payload is not determined by the channel bandwidth or data format (digital or analog). With the exception of problems related to amplification (e.g., EDFAs) and other fiber impairments, transparent optical networks are suitable for transmission of quantum states. In these networks, photonics is used not only for transmission but also for networking functions such as multiplexing, switching and wavelength add/drop. This allows the establishment of reconfigurable end-to-end wavelength 'light paths' through the network, without any O-E or E-O conversions. Similarly, one can transmit classical data on some wavelengths, and quantum signals on others. Carrying classical and quantum signals over the same fiber is referred to as 'coexistence.' This has been successfully demonstrated by combining conventional classical telecom channels with QKD signals over access, metro area, and even (segments of) long-haul links. Even if classical data is not transmitted in the same fiber as quantum signals, coexistence is very often necessary as classical-level signals for synchronization and stabilization of the quantum light path are needed. However, the extremely large optical power mismatch between classical and quantum signals requires an understanding of the key impairments and very careful designs. In addition, applications such as quantum computing and sensing are likely to require far higher fidelities than QKD, which typically operates at quantum bit error rates (QBER) approximately a few percent. Coexistence will be a critical research issue and a practical consideration for quantum networking architectures. Finally, while traditional WDM networks have relied on fixed-width wavelength channels, newer approaches such as FlexGrid could provide greater flexibility for reconfiguring and reallocating the optical spectrum according to changing service demands, or for adjusting guard bands between highly power-mismatched channels.

## 2.3 DOE's Quantum Networks – Beyond Quantum Point-to-Point Links

The DOE is a complex science organization with national and international computing resources, science facilities, and research teams. Its distributed nature made classical networking a critical component of its scientific infrastructure, and this should not be different for quantum networks. From this perspective, DOE's emerging end-to-end quantum networking infrastructure should be seen as a collection of autonomous quantum networks as shown in Figure 5. It consists of Q-LANs in various laboratories and university campuses, Quantum Metropolitan Area Networks (Q-MANs) serving regions, all of which are linked together by Q-WANs analogous to DOE's current ESnet backbone networks. Such segmentation simplifies network management and enables different segments to evolve independently in a scalable way. While this approach of loosely interconnected autonomous network segments worked well for classical networking, it is not clear how it will work for quantum applications that require tight synchronization. While transparent optical networks have many attractive features, they typically do not include wavelength conversion, which may be useful for quantum networks.

In the DOE, networks serve a support role. In short, they exist to serve the needs of DOE users. This means that networking efforts must support DOE missions. At the same time, quantum networks must complement and coexist with the DOE science complex consisting of supercomputers, science instruments, analytics facilities, and networks (ESnet and site networks). Consequently, the design space spans not only the quantum domain but also conventional networking domains. In practice, quantum applications usually require a combination of quantum and classical information transmission, for example to connect, control, and process information from distributed quantum processors or sensors. In addition, it appears that several different types of quantum networks will be necessary for the DOE. Some might be achievable in the not-too-distant future, e.g., interconnecting quantum computing devices within a lab using a Q-LAN. This might be necessary, for example, when one builds a quantum computer larger than can be handled by the cooling capacity of a single dilution refrigerator. Others may require advanced technology that does not currently exist, such as quantum repeaters for transmission over longer distances using Q-WANs.

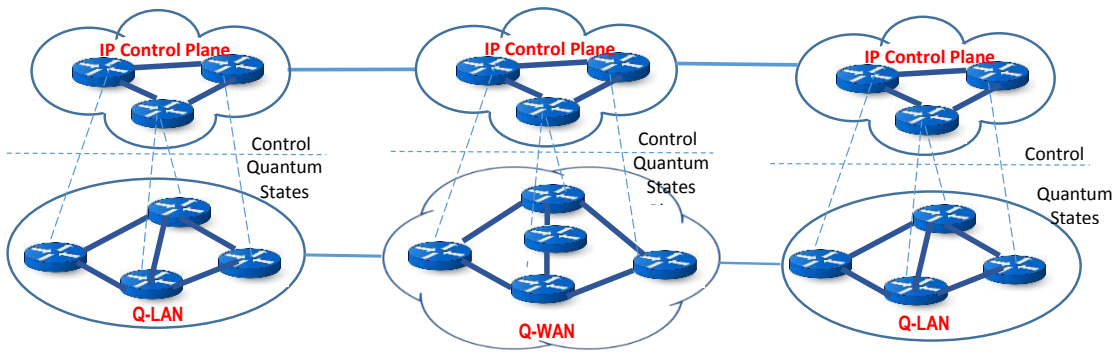

**Figure 5**: End-to-End Transparent Optical Networks with Digital Control Plane

## 2.4 Quantum Network Architecture

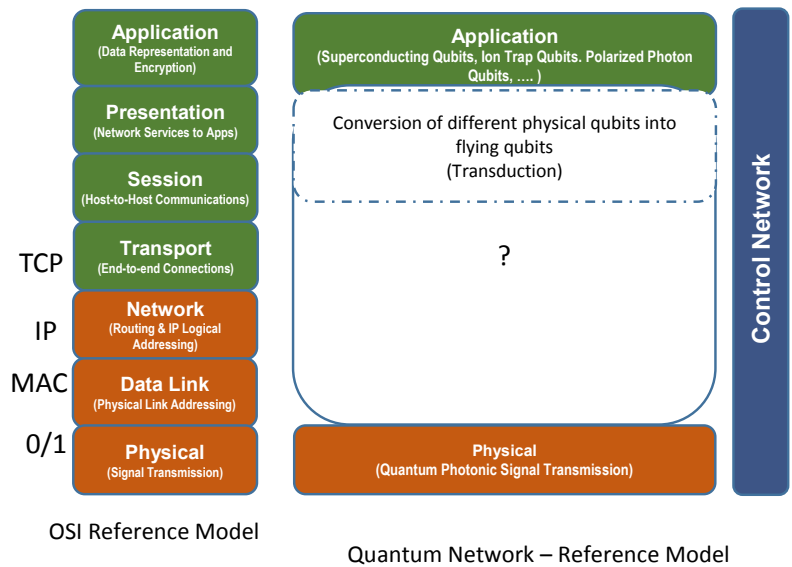

**Figure 6**: Quantum network reference architecture compared with classical network OSI Model

Communication networks are complex dynamical infrastructures. Designing and operating these systems is challenging, especially if they are to be scalable, easy to operate and manage, and accommodate multi-vendor subsystems. These challenges have been successfully addressed in the design of classical networks thanks to the adoption of the Open Systems Interconnection (OSI) layer model for network architecture (shown in Figure 6). Network layering allows complex end-to-end communications tasks to be decomposed into logically manageable groups of services and protocols that communicate with other layers through well-defined interfaces. Given its popularity and success, a fundamental question that emerges in the nascent quantum network community is "What lessons can designers of quantum networks learn from the success of the internet as applied to classical networks?" Quantum information carried over quantum networks is characterized by unique features such as entanglement and constraints such as no-cloning.

These features and constraints may prevent a direct mapping of classical network layers to the quantum network layers. The universal functions and services in a given layer in a quantum model are still unknown as shown in the reference architecture shown in Figure 6 that compares transparent optical network architecture to the OSI model used for classical communications. For example, the current internet is designed around the concept of Internet Protocol (IP) packets, which forms the basis of reliable connections, network addressing, network flow and congestion control functions (the OSI reference model is shown in Figure 6). Currently there is no notion of a "Quantum Packet," - a photonic quantum state along with appropriate headers that function as a single data unit which traverses the quantum network. Experimental quantum networks to date have focused primarily on physical layer photonic transmission. Additional investigation is required to formulate a framework that captures and organizes the functions, services, and protocols needed for layered quantum network architectures. Critical questions that must answered are:

- What are the basic services that quantum networks should provide to DOE scientists and, in the future, to the general QIS user communities?
- There are currently several physical qubit implementations, such as trapped ions, superconducting qubits, and polarized photons. Others are expected to be invented in the future. Should a quantum network be designed to support all these different qubits types or a specific type amenable to optical networks?
- What quantum states need to be shared by the network, e.g., the encoding: qubit vs. qudit vs. continuous variable, the type of entanglement, two-particle or multi-particle? What are the system tradeoffs for using complex quantum states (e.g., hyper-entanglement and qudits) to improve bandwidth utilization and throughput?
- Given that quantum photonic states cannot be processed while traversing a network as is done classically, how should network traffic engineering and quality of service be supported?

Central to the above questions is whether or not a qubit, and if so what type, should be the fundamental building block in open science quantum networks. The answers to these questions are critical in developing a scalable quantum network architecture that will not only support DOE missions but could be extended and made applicable to the emerging quantum internet development effort. Requirements for these networks, and the range of options for their successful implementation, is at present very poorly understood, and requires research and investigation.

## 2.5 Quantum Network Control

The ability to control, optimize, and recover from failures are critical in the design and operation of large-scale networks. Achieving these capabilities typically requires networks to carry network management and control information in addition to user data. There are two distinct methods to implement it: a) in-band control, if control and management information is carried on the same channel as the users' data and b) out-of-band, if it is carried on a separate channel. It is clear that quantum networks will need out-of-band control and signaling such as shown in Figure 5, since any attempt to read and process control information carried in the quantum channel will destroy its content. Out-of-band signals are common in telecommunications systems where the information carrying signals cannot be internally processed because of time constraints or because of lack of processing technologies. In classical voice telecommunications, this is known as Signaling System 7 (SS7) and in transparent optical systems, it is known as Generalized Multiple Protocol Label Switching (GMPLS). In both cases, the information carrying signals are analog. The case for an out-of-band control network for quantum communications networks is compelling because, even if quantum processors were available to support in-band control, their usage will be limited since any attempt to read quantum information will irrevocably lose some quantum information. GMPLS is an extension of Multi-Protocol Lambda Switching (MPLS) signaling and internet routing protocols to provide a scalable, interoperable, distributed control plane, which is applicable to multiple optical and digital network technologies such as optical cross connects, photonic switches, IP routers, Asynchronous Transfer Mode (ATM) switches, Synchronous Optical Networking (SONET) and DWDM systems. GMPLS flexibly accommodates both digital and optical network control signaling, making it attractive for transparent optical quantum-classical coexistence networks.

Quantum networks, despite their peculiar nature and sensitive requirements, will not exist in isolation. Quantum traffic will coexist with other types of network traffic in the same network infrastructures, as has been demonstrated in many QKD trials [15, 21, 24, 25]. It is anticipated that the early deployment of quantum networks will need to coexist with classical IP networks. For these two types of traffic to coexist in the same fiber and share the same network resources, several challenging technical issues have to be resolved, as they have conflicting requirements. The type of bandwidth sharing is at the wavelength level. Further research will be needed to support classical IP traffic and quantum signal traffic. Fiber sharing will occur at the optical spectrum levels using DWDM or CWDM techniques that partition the available optical spectrum into grids or channels as shown in Figure 4. Wavelength sharing could be enabled by using some other type of multiplexing, for example time-division multiplexing. However, quantum traffic cannot be part of the same spectrum that is amplified (see Figure 4) or processed by O-E-O. In addition, there are different requirements for classical and quantum wavelength conversion. The design space of a resulting hybrid quantum-classical coexistence network will be complex and unprecedented, since it not only involves the development of quantum networks but also their interfaces with conventional counterparts. In particular, new mechanisms and protocols are needed to support classical and quantum data and control flows in both parts with a particular emphasis on crossovers, including monitoring and control of quantum components using classical network management, and exploiting the security and other capabilities that can be provided by quantum to enhance the classical network. These considerations lead to an expanded design space of quantum and classical parameters, and their interactions. A major challenge in realizing the coexistence of quantum and classical information is how to reduce the crosstalk from strong classical data channels into the quantum channels. Due to intrinsic properties of telecom fiber, noise photons can be generated from various processes, such as Raman scattering, four-wave mixing, etc. In the context of QKD, several solutions have been established. In discrete variable QKD based on single photon detection, heavy filtering in the frequency domain, the time domain, or both frequency and time are typically required [25-28]. In continuous variable QKD based on coherent detection, thanks to the intrinsic filtering function of the local oscillator, external filtering is typically unnecessary [29, 30].

## 2.6 Challenges and Opportunities

**Challenges**

- What services can be efficiently supported by these architectures?
- What are the requirements of each service in terms of the types of quantum states, throughput, and fidelity that are needed?
- Can the network support a diversity of multiple simultaneous services? Should it support delivery of quantum states with different fidelities as required by different services? Alternatively, should there be a 'best-effort' baseline fidelity, and if so, what is required at the endpoints to meet the service requirements?
- What combinations of physical topologies, distances, and numbers of endpoints need to be supported?
- What specific quantum technologies are required for these architectures, and when will they be ready?
- What services should the network support natively, and which should be provided by applications layered over the basic network services?
- What are the vulnerabilities of each conceptual architecture, when attacked?

## Opportunities (Short-Term, 2 - 5 years)

Some of the key DOE quantum networking research issues can be addressed immediately, with results available in the next 2-5 years, while others may require decade-long efforts. The remainder of this section identifies key research issues.

- Development of performance models and simulations, and experimental efforts to leverage existing technologies
- Exploration of novel classical-quantum coexistence network architectures, protocols, and services
- Analysis and testing of quantum-classical coexistence networks
- Exploration and testing of switching mechanisms such as dynamic circuit switching, burst-switching, and packet-switching for quantum networks

## Opportunities (Long-Term 5- 10 years)

Longer-term efforts will bear fruit when underlying technologies become available (e.g., functioning quantum repeaters). These efforts will rely on a combination of basic and applied research and will need steady funding over many years before their results fully pay off. These efforts, though long term, will have very high payoffs as they succeed. In the near term, systems-level analyses will provide critical insights into prioritizing goals for the longer-term research. There will also be benefits from significant, ongoing interactions with shorter-term research objectives, so that researchers in each type of effort may learn from each other. Key research topics include:

- What are efficient protocols and topologies for a realistic quantum internet once high-throughput sources, repeaters, etc., become available?
- What are the requirements for a quantum control plane?
- Will the technologies developed for the quantum internet have implications for the classical internet?

# 3 Quantum Network Devices and Subsystems

Moving from single quantum systems to interconnected quantum networks brings new problems associated with loss (distribution loss), delay (memory time), distance (synchronization and coherence), and heterogeneity (interfacing different technologies). In particular, loss is a ubiquitous issue that ranges from the mundane but important application of optimal mode-matching to complex system optimizations and materials science questions. Especially in complex systems, fractions of a dB are important. This high sensitivity to loss largely separates the realization of complex quantum systems from complex classical systems.

Multiplexing is an often-pursued approach to build quasi-deterministic quantum signals from probabilistic signals. It usually also requires highly optimized classical components, especially with regard to loss. Other technologies that are compatible with ultra-low losses and large numbers of modes should be investigated.

In general, quantum device makers could use a better understanding of how devices and their design choices fit into larger quantum networking systems. Perhaps a tiered rating system of goals that would enable different applications would be helpful. Nevertheless, specifying different performance metrics that would make a component technology helpful, important, or ubiquitous would be useful to the development process. Such an understanding requires analysis of case studies of the quantum network applications under different assumptions.

## 3.1 Quantum Encodings

Quantum networking requires the creation, encoding, transport, processing, transduction, and measurement of photonic quantum states. For telecom network applications utilizing the existing fiber-optic infrastructure, the photonic quantum states could be in the O, C, U, or L bands. Future fiber types, such as air core or chalcogenide fibers could enable other spectral transmission bands. The quantum information may be encoded in any photonic degree of freedom, and these encodings follow two differing approaches: discrete variable (DV) encoding, such as a qubit defined by two orthogonal polarizations, and continuous variable (CV) encoding, such as a particular phase from a broad continuum of possibilities. Experimental effort to date has mostly focused on discrete approaches realizing qubits.

However, while the qubit-based DV quantum computing approach has a longer history, the qumode-based CV quantum computing approach has recently drawn more attention [31]. Thus, future quantum communications networks will likely need to support transmission of both CV and DV quantum states to network future quantum computers. In a specific area of quantum communication—quantum key distribution, both DV and CV protocols are well developed. CV-QKD protocols based on coherent detection are especially appealing for their compatibility with standard telecom devices. For example, the well-known Gaussian-Modulated Coherent State (GMCS) QKD protocol [32] can be implemented using conventional attenuated laser sources (instead of single photon sources) and compact, high-efficiency balanced photodiodes working at room temperature (instead of single photon detectors working at low temperature). CV-QKD's use of coherent receivers not only makes it an attractive candidate for chip-size implementation [33], but also greatly enhances its resilience to broadband noise photons when classical communications and CV-quantum signals coexist in the same optical fiber. The improved noise tolerance is due to the intrinsic filtering function of the local oscillator [29, 30]. The above advantages are likely preserved for other CV quantum communication protocols, though it remains an open research topic. Of course, the CV quantum network approach comes with its own challenges. For example, in contrast to a DV quantum repeater, which has been extensively studied for over 20 years, the CV quantum repeater is still in its infancy [34, 35]. Much more work needs to be done on CV quantum repeaters to compare their strengths and weaknesses to those based upon DV approaches. Considering that a future quantum internet will likely be heterogeneous, it is also important that research develop interfaces between DV and CV quantum devices and hybridization promises some advantages [36-38].

## 3.2 Quantum Network Devices

Although they will draw from lessons learned from transparent optical networking, novel core quantum networking functionalities will need to be developed to work together in a harmonious system. As quantum signals need more careful treatment than classical signals, some very different design choices will need to be made. The biggest difference is the need for quantum repeaters to replace analogous classical technology to reshape, retime, and re-amplify classical signals. Other differences will be manifest in the need to control out-of-quantum band signals for transmission impairment monitoring, such as polarization changes from fiber birefringence. Further capabilities will be needed to route and switch quantum photonic signals much more quickly, with lower noise, and with higher efficiency than is needed in classical optical networks. Quantum photonic sources come in a wide variety of configurations but are largely based on two principles, either emission of an excited quantum system, such as a quantum dot, or the nondeterministic process of nonlinear optical down conversion. Some specific applications such as decoy-state quantum key distribution can utilize weak coherent states, such as produced by an attenuated laser. While useful for particular quantum key distribution protocols, weak coherent light sources are generally not as useful for more generic DV quantum networking due to the statistical emission of multiple photons. Parametric down conversion sources typically emit photons at much lower probability than weak coherent sources, to reduce the likelihood of emission of extra photons. Sources are normally configured to produce either a single photon or a pair of photons, and for the latter, they are typically entangled. One- and two-photon sources normally utilize DV encodings. For fiber optical networks, the most common qubit encodings utilize polarization and phase (e.g., between time bins) degrees of freedom. Other less common encodings utilize d-level systems, such as the phases between d time bins, and are called qudits.

Qudits are largely utilized for QKD applications due to their increased noise tolerance [39]. If multi-core fibers become common, perhaps spatial encoding will become applicable to fiber optical quantum networks.

However, less common for telecom network applications, some sources of squeezed entanglement have been investigated. Loss is problematic for all quantum light sources, reducing throughput and making sources, which even if in principle deterministic, nondeterministic. Although they are nondeterministic, most quantum communications system demonstrations utilize down conversion photon sources due to the high quality of produced photons, or weak coherent pulses. Creation of deterministic single-photon and two-photon sources remains an experimental challenge. The properties of generated photons typically must be precisely controlled to perform quantum interference with other photons or to interface with other systems, such as quantum memories. Typically, the goal is to produce photons, which are indistinguishable, to enable high-quality interference, an important function in many quantum operations.

### 3.2.1 Transduction Devices

Transduction could be for photonic quantum frequency translation or conversion between a photonic mode and some other non-photonic excitation, such as a motional degree of freedom. Generally, it is desirable to be able to convert non-telecom frequencies to telecom frequencies for optical transmission as well as from telecom to non-telecom at a network node. Quantum transduction between very distant electromagnetic frequencies, such as shifting from an optical carrier suitable for distribution over fiber to a microwave suitable for superconducting qubits, remains an experimental challenge. Such a capability would enable one to use heterogeneous qubit types for specialized quantum functions. Quantum transduction would likely be a critical technology and there has been substantial progress but, the conversion efficiency, noise, and/or bandwidth characteristics of current systems are not yet adequate. Substantial work is required on this front including advancements in the theory, design, and implementation.

### 3.2.2 Quantum Frequency Conversion

Quantum frequency conversion can be carried out to shift the wavelength of a photonic qubit through an electro-optic or nonlinear optical interaction (with a $\chi^{(2)}$ or $\chi^{(3)}$ nonlinear material). These techniques require specific configurations, as not all classical frequency conversion methods have low enough noise to preserve quantum states. Historically, due to the poor performance of single photon detectors for telecom wavelengths, quantum frequency conversion from telecom to visible light via $\chi^{(2)}$ nonlinearities has been studied extensively in an effort to translate photons for detection by silicon avalanche photodiodes. It is also important to be able to convert visible photons to wavelengths suitable for optical fiber transmission. Many configurations have been tried and some commercial products are even offered. However, while realizing efficient, low noise conversion has been achieved in many experiments, such devices are not yet mature enough that they are widely available for use in larger quantum networking systems. Quantum frequency conversion is further complicated by requirements to match very different spectral characteristics for different parts of the quantum network, for example, an optical pulse optimized for optical fiber transmission with spectral characteristics optimized to interface with a quantum memory.

### 3.2.3 Quantum Repeaters and Routers

To counter the impact of loss and the inapplicability of classical techniques, various proposals have been made for quantum repeaters. A quantum repeater can in principle be used to avoid the impact of exponential loss, enabling continental-scale quantum networks. However, experimental progress towards realizing the theoretical promise has been slow. The earliest concepts for quantum repeaters relied upon the distribution of two-qubit photonic entanglement, after which the entanglement could be used to teleport the state of some qubit. To establish long-distance entanglement requires several steps. One proposal breaks up the overall link length into shorter sections where entanglement would be distributed pairwise between adjacent nodes to couple nearest neighbors. Then each node, except for the start and the end, would have an entangled photon from each neighbor. Next, interior nodes would perform a joint "Bell state" measurement on its two photons in an operation called "entanglement swapping" to establish entanglement over the long-distance link. As photonic transmission is probabilistic along each link, various proposals adopted the use of quantum non-demolition measurements to signal the successful arrival of a photon and load it into a quantum memory, where it would wait for the arrival of a similar photon entangled with the other neighboring node. In practice, quantum non-demolition and quantum memory steps add noise. To improve the success rates, various proposals theorized sending many entangled pairs simultaneously. While not all of the photons would make it to each node, those that did could be combined into a fewer number of entangled links with lower error by using some post transmission quantum processing called distillation, concentration, or purification. This processing requires high quality two-qubit gates and two-way classical communications. The arrival of fault tolerant quantum gates and memories could improve some aspects of performance. Other similar svariations have been proposed. More recently, quantum repeater protocols, which require only one-way classical communications, were proposed. These types of quantum repeaters encode a logical qubit into many photons, which are transmitted to the next quantum repeater where quantum error correction is performed to fix errors due to photon loss or other experimental imperfections. The quantum error correction-based repeater can transmit fault-tolerant quantum data in real-time, so it is ideal for interconnecting quantum processors, either from chip to chip, or over longer distances. This repeater type is effectively a small purpose-built quantum computer.

### 3.2.4 Quantum State Multiplexers/De-multiplexers

These devices enable quantum states to be groomed (aggregated) into a common payload to share a common quantum channel. The intent is to aggregate low-speed traffic to form a bigger payload to be carried in network channels with higher bandwidth. For this to work efficiently, the network traffic must have an address or a label that will enable individual flow of packets to be identified when the payload is disaggregated at the destinations. In classical networks, IP addresses and flow labels are used to provide these functions. Quantum signals may require classical herald signals, which carry routing information. Alternately, quantum channel assignments (e.g., wavelength, time slot) can be made and managed exogenously.

## 3.3 Network Design: Performance Modeling and Simulation

Modeling and performance analysis are important, both in the design phase to evaluate and compare the merits of a variety of quantum network protocols and architectures, as well as for real-time performance analysis and troubleshooting after the network is built. Simulations will be needed to study network properties including quantum state and entanglement throughput, latency, scalability, reliability, and availability. Since generic quantum systems consisting of even hundreds of qubits cannot be fully simulated on classical computers, ways to effectively employ reduced models are required. Some methods already exist that will clearly be useful, such as Monte Carlo simulations of systems whose operations only include Clifford gates. Some questions, for example, the extent to which various protocols are able to avoid bottlenecks, will be able to be addressed by purely classical simulations, using methods similar to those already developed by the classical networking community. Other questions, pertaining to the physical layer, will require simulation of the dynamics of optical channels and their interaction with the systems that comprise sending and receiving circuits. Such simulations will likely require the use of much more sophisticated methods such as matrix-product-state methods or tensor-network methods.

The question of simulating networks to evaluate performance also raises the question of what metrics are necessary to characterize their performance (e.g., the Tangle, Fidelity, entropy, quantum channel capacity, etc.). Moreover, what measurements must be made on the system in order to capture real-time performance and diagnose hardware issues or bottlenecks? To support any new metrics necessary to characterize quantum network performance, specifications of those metrics and the underlying measurement methods to acquire the parameters are needed. New instrumentation may need to be developed to perform the measurements crucial to compute those metrics. This instrumentation can also be used to diagnose the health of the network and help pinpoint problems.

Last, queueing theory has played an integral part in developing an understanding of and tools for the evaluation of performance of classical networks; what is the analog needed for modeling and evaluation of quantum networks? Simulation and modeling research questions include:

- What kinds of simulation methods and tools will be required for exploring the performance of network architectures?
- What kinds of metrics will be required to capture the performance of quantum networks?
- What kinds of quantum many-body effects will arise in quantum networks that impact the performance of the networks?
- What kinds of mathematical tools will be required to analyze the behavior of quantum networks (e.g., percolation theory and queueing theory for classical networks)?
- What kinds of simulation tools will be required to evaluate the performance of networks in real time, to determine, e.g., when nodes have failed, necessitating re-routing?
- What kinds of network element measurements need to be made to understand the functioning, efficiency, and sources of errors in a network?
- What characteristics in existing network elements are inconsequential (or manageable) for classical signals, but relevant for quantum ones? e.g., nonlinear properties of optical fiber, Brillouin scattering, Raman scattering.
- How should experimental testbeds be designed and utilized to complement simulation tools to support designs?

The above questions regarding simulation tools and diagnostic methods will need to be informed by the structures of the networks that are to be modeled and the kinds of questions to be asked. In the short term, it is expected that networks without quantum repeaters for various applications can be proposed now, such as distance-limited communications between quantum computers or quantum sensors. In the long term, long-distance networks with repeaters, or their alternatives, can be proposed and modeled for those same applications, as well as other applications that will emerge.

## 3.4 Challenges and Opportunities

**Challenges**

- Unlike classical networks where a certain loss budget is tolerable with the addition of amplifiers, excess optical loss remains a key issue, which requires significant device and subsystem engineering to solve.
- Creation of deterministic non-classical light such as single photon, two photon, and multi-photon states remains an experimental challenge, even if these states are not entangled.
- In addition to loss errors, it is not clear how to correct other operational errors, which occur at the device and subsystem level.
- Transduction of quantum information from fixed qubit technologies to flying qubit technologies and back is not fully developed.
- Quantum repeaters will be required to build long-distance networks yet do not appear poised to break even with simple direct transmission in the near term.

> **Opportunities (Short-Term, 2 - 5 years)**
>
> - Some quantum networks should prove simpler to simulate than other quantum systems due to the more limited interaction types
> - Quantum frequency conversion is poised to become a commercially available technology
> - New forms of quantum frequency grooming via electro optic techniques could see dramatic efficiency improvement by moving to integrated optical platforms
> - Continuous variable approaches may provide more attractive methods towards realizing quantum repeaters, yet have been researched far less theoretically than discrete approaches
>
> **Opportunities (Long-Term 5- 10 years)**
>
> - Error corrected quantum memories could prove useful in building new types of quantum repeater networks
> - Mature transduction between static and flying qubits promises to enable exponential quantum computing capability by linking distributed resources
> - New types of efficient multi-photon sources could enable efficient photonic quantum repeaters

## 4 Network Operations and Management

Quantum networks are complex, challenging engineered systems that require sophisticated solutions for their operations and control, with many of those solutions yet to be developed. Indeed, many of the control plane technologies in use in modern classical networks are not suitable for the quantum data plane that cannot be subjected to O-E-O conversion, as discussed above. Quantum network management and operation will be particularly challenging due to the quantum nature embedded in the control plane and/or the data plane. The task is further complicated by the need for quantum networks to co-exist with conventional networks. In addition, monitoring of quantum networks requires measurements of complex conventional and quantum signals, along with inferences and analytics to distill knowledge and make control decisions.

The basic functions of quantum network management and operation will include most of the same elements that are found in classical network management [40]: performance management, fault management, configuration management, security management, and accounting management. Quantum networks, regardless of centralized or distributed management planes, will likely require all these functions, as well as additional functions that may be required to handle the complexities of quantum information. In addition, a quantum network will exist within an ecosystem that also includes a classical network. Thus, the overall strategy must encompass operation and control functions for both networks. A potential starting point for the development of quantum network management and operation techniques is the extensive knowledge base associated with ESnet. Several advanced tools are in daily use in ESnet, including a leading-edge control plane, continuous fine-grained network monitoring, detailed models of expected network behavior, and trouble-shooting aids. These sophisticated management techniques, however, will need to be significantly extended to control the large system of interlinked quantum devices that will comprise the quantum network. Some of the more significant challenges for the management and operation of a quantum network are detailed below.

## 4.1 Network Monitoring and Performance Management

Networks typically consist of many components that work together to deliver the network service. The goal is to setup and enable links and then to monitor them to estimate the quality of service. These components, the routers, switches, amplifiers, modems and more need to work in a coordinated way to achieve the performance level required. Typically, network control systems work by monitoring and measuring the current state of all the components, receiving requests for services, and then issuing commands to the different components to bring them in the desired state. These control systems often rely on models of components and network architectures to optimize the service delivery and to mitigate errors and outages. Several aspects of this type of approach will be challenging when applied to quantum networks.

First, because quantum states cannot be measured without being altered, the quantum payloads cannot be used to directly monitor the network condition, and so alternative approaches will need to be developed to infer the health of the network. For example, it should be possible to probe many of the components with classical fields to learn something about their operating conditions. For components for which this is not practical, it may be necessary to employ non-computational ancilla photons dedicated specifically for network monitoring activities. This could be done either independently of the quantum payload or as part of a larger quantum state that includes "network monitoring" ancilla (in a manner analogous to quantum error correction).

A second challenge to performance management is that a complete model of a large quantum network is likely to be impractical. While it is generally possible to model small quantum devices, the complexity grows exponentially as they grow larger, making it impractical to model large-scale systems. Models of individual components such as routers, switches, and even quantum repeaters, although challenging, should be achievable. In addition, it should be possible to construct a larger network model from abstractions of these component models. However, these larger models are not likely to accurately predict the behavior of large quantum states distributed across the network. A similar problem is encountered in quantum computing, where it is not possible to completely model devices of even modest size. A promising approach in this case is to model the device as a black box and model its behavior, rather than its complete quantum state. A similar approach might be adopted for quantum networks.

## 4.2 Quantum Network Traffic Control

Quantum networks eventually will evolve to wide-area networks consisting of multiple autonomous systems, built with heterogeneous vendor devices and perhaps with different technologies. Coordinating the spectrum allocation of multiple autonomous quantum networks to achieve end-to-end quantum channel provisioning with high spectral efficiency and quality of transmission, along with security guarantees, is not trivial. Whether distributed or centralized control and management is more suitable for multi-domain quantum networks remains an open question. While distributed systems are inherently more scalable, they may suffer from slow convergence and low resource efficiency due to the lack of coordination among domain managers. On the other hand, centralized schemes may improve the effectiveness of multi-domain quantum networks by introducing a global resource orchestrator that dictates the operations of domain managers. For different multi-domain quantum network architectures, efficient inter-domain traffic-steering algorithms will need to be developed, specifying, for example, which information each domain should expose and how an end-to-end quantum connection should be determined. In a distributed system, the source quantum domain may decide the domain sequence first based on its multi-domain connectivity (e.g., obtained through the Border Gateway Protocol) and set up the inter-domain connection domain by domain. On the other hand, the multi-domain orchestrator in a centralized system could collect multi-domain abstractions (e.g., full-mesh abstraction) and calculate inter-domain connections by performing global traffic engineering.

Quantum network traffic control strategies must also consider state preservation and synchronization issues. To enable photonic quantum state transmission over macroscopic distances, the transport channel must preserve the state's coherence or superposition. It is likely unavoidable that a quantum state injected into a link will change as it propagates. This is not a problem, as long as these changes can be undone with intervention of network devices. For point-to-point links and "Alice-Bob" applications this is conceptually straightforward. However, the task becomes much more complicated for a large-scale network with, perhaps, arbitrary connections. In a network with arbitrary connection and traffic, different types of quantum states from many users may be co- or counter-propagating along common links. Thus, these links cannot cater to any individual, but must operate to serve all network users.

This poses the question as to whether all stabilization should be pushed to the network edges. Pushing stabilization and synchronization to the end users is attractive, as the desired quality may be application dependent. However, this seems problematic, since quantum states will need to be routed through many nodes. To that end, it may be necessary to establish concrete bounds on the degrees of freedom allowed (e.g., temporal extent of a state, and frequency bandwidths).

## 4.3 Quantum Network Security

There are unique and extremely important security questions involved in the reliable, trustworthy operation and control of quantum resources to support the DOE's quantum computation and quantum sensing efforts. Within the framework of coexistence infrastructures, security vulnerabilities of conventional networks carryover, and those of (newer) quantum components need to be explored and addressed. Furthermore, novel crossover vulnerabilities may potentially exist, wherein one modality may be exploited to compromise the other. Indeed, these aspects must be addressed from the beginning as an integral part of the design and analysis.

Although there has been considerable investigation of the security properties of QKD systems, the security of a larger quantum network has not received as much attention and there is much that is unknown. For example, the relationship between point-to-point entanglement and security is well understood. However, interesting new possibilities (and potential vulnerabilities) arise when entanglement is distributed across multiple locations. Moreover, security vulnerabilities could propagate both ways in infrastructures with co-existing quantum and conventional networks; these inter-dependencies need to be examined and explicitly addressed to ensure the overall security of the infrastructure.

On the positive side, quantum network capabilities could enhance the security posture of the conventional network and the control plane implemented on it. A quantum channel that coexists with classical channels in a common network could provide security enhancement to the network as a whole. The quantum channel would be easier to monitor for adversarial behavior, and the quantum channel can securely transfer important control information to the node management subsystem.

## 4.4 Challenges and Opportunities

**Challenges**

- It is not clear what, if any, existing classical network simulations are extensible to and compatible with quantum networks.
- Due to the quantum nature of flying qubits, it is expected that simulation of distributed quantum resources will be more difficult than classical bits.
- The conventional and quantum measurements to extract the critical knowledge and actionable information for operation and control of the multi-site infrastructure are complex.

**Opportunities (Short-Term, 2 - 5 years)**

- ESnet fiber infrastructure together with national lab site fiber infrastructure are primed to host research to make progress on the quantum networking challenges described in this report.

**Opportunities (Long-Term 5- 10 years)**

- A multi-function quantum network testbed that integrates quantum computing and quantum sensor devices would provide a foundation for a multi-science mission quantum network.
- A quantum-classical coexistence ESnet for the DOE science complex consisting of various national laboratory sites will enable a leap in the nation's capability by connecting emerging quantum devices.

# 5 Path Forward

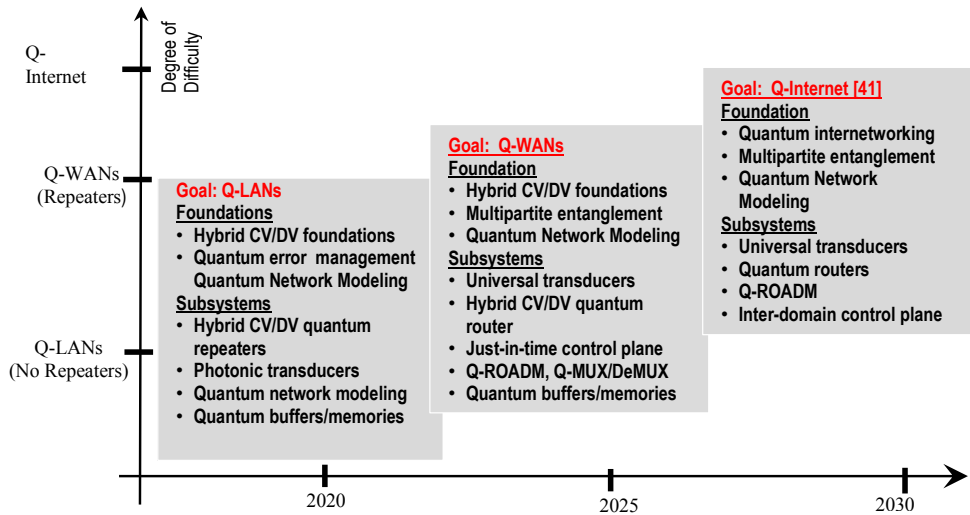

Figure 7: Transparent optical Quantum networks evolution in the open science

Quantum communication networks are a nascent technology. As a society, we have come to expect more out of computing, especially networking, and cannot imagine a day without it. We want networks that are compatible with our computing needs, faster, secure, and accessible anywhere and anytime. The emergence of a quantum-computing paradigm that is radically different and incompatible with the current mode of communications presents a unique challenge. Quantum networks have a solid and well-documented quantum-mechanical theoretical foundation but not much is known about translating it into a practical implementation for most of the science applications listed in Section 1.3. This is due, in part, to the multi-disciplinary effort that is required. The US telecommunications industry is very economically competitive and needs a compelling market opportunity as well as a practical way to evolve existing networks to add these types of quantum communications-based services. As a result, they are not in a position to drive these changes or the enabling early-stage research, which is why government investment is critical. Although much progress has been made on quantum networking for QKD, and the ability to carry QKD over existing networks, these approaches are not necessarily generalizable to support the much more challenging demands of a broader range of quantum applications.

There is some sentiment in the community that progress on wide area quantum networks depends on the progress of quantum repeaters to extend quantum transmission beyond a few hundreds of kilometers. While this is true, progress towards the realization of quantum networks can be grouped into three major categories corresponding to their geographical reach:

- Category 1. Repeater-less Q-LANs and Q-MANs
  These are generally networks within campuses, national laboratories, and metropolitan areas where a quantum link has a maximum span of approximately 150 km. At this maximum distance, most transmitted photons will be lost, and the supported applications may be limited. Some applications are supported without quantum repeaters, though their inclusion is likely to improve application performance. For this category of quantum networks, potential open problems include the following: a) packaging several quantum states into a single addressable entity analogous to IP packets that is routable across a quantum network; b) new fast, low-loss, and low-noise quantum network switching in analog to the classical techniques of circuit switching, packet switching, or burst switching; c) quantum state traffic grooming; d) error management before quantum error correction is available, and e) quantum resource (e.g., entanglement and non-classical light) management.

- Category 2. Q-WANs
  At the core of this category of quantum networks are quantum repeaters and associated devices such as routers, non-blocking optical switches, quantum memories, reconfigurable ADD/DROP multiplexers, and frequency converters, which extend the reach of quantum links beyond 150 km.

- Category 3. Quantum Internet
  These networks have the characteristics of coupling multiple autonomous network domains requiring coordination of different network policies, quality of entanglement (QoE), and synchronization of control planes. These three categories of quantum networks require various challenges to be solved and improved upon to realize quantum network devices such as quantum buffers and memories, quantum repeaters and routers, and crosscutting activities such as quantum network performance modeling, quantum networks standards and metrology, and multipartite entanglement establishment. These capabilities will require a long-term development effort.

## 5.1 Inter-Agency Collaborations

Quantum communications networks are of interest to federal agencies, which depend on advances in information technologies to execute critical aspects of their mission. These agencies, like DOE, not only have to use the state-of-art information technologies available but also have to develop new technologies, such as quantum networks, if they are not commercially available. Given that the missions of these agencies are complementary, it is highly likely that they will explore and develop complementary quantum communications networking technologies. This provides the opportunity for collaboration and sharing of ideas that could be mutually beneficial. For example, the quantum network metrology and standards efforts at the National Institute of Standards and Technology (NIST) is a critical component needed to ensure interoperability of components and protocols in the realization of quantum internetworking across different administrative network domains. The National Aeronautics and Space Administration's (NASA) effort on space-based quantum communications networks is another ongoing activity that will innovate on different aspects of quantum networks that can be leveraged by other agencies. Finally, the Department of Defense's research labs are developing quantum technologies, which may be useful for science applications. Inter-agency collaboration will greatly strengthen the impact of developed quantum networks.

## 5.2 Industry, Academia, National Laboratory Collaboration

Quantum networks, like many other innovations, which originate from basic research in academia and national labs, face technology transfer challenges despite their overwhelming potential to increase the nation's capabilities and benefit its society. Even though HPC and high-performance optical networks have been developed in concert to improve computing capability, major information technology providers investing in quantum computing, such as Google, IBM, and Microsoft, and others in the telecommunications sector still view quantum networks as a high-risk effort. This means that the government must play an active role in prioritizing and matching investments from the private sector. As a neutral actor, the government could also facilitate the development of standards that will be critical in building inter-operable subsystems critical for quantum telecommunications.

## 5.3 Quantum Networks Standards and Metrology

The success and ubiquity of the modern internet is largely due to the coordination of industry, professional, national, and international networking standardization efforts. Network standardization helps ensure that communications equipment, protocols, and software stacks from different vendors can be integrated in a way that enables them to interoperate seamlessly.

Due to the complex nature of quantum systems, standardization efforts will be critical in the development and testing of the first generation of quantum networks. Potential issues to be addressed include specification of performance metrics, noise and error level specifications, the development of a reference architecture, and building a common terminology and taxonomy among the multidisciplinary community that is needed to solve the challenges outlined in this report.

## 5.4 Challenges and Opportunities

**Challenges**

- Room temperature quantum memories and buffers are essential for scalable quantum networks and real-word science applications.
- Quantum networking is a complex, high-risk but high pay-off technology that will take many years of investment to deploy practical systems for DOE.
- A quantum workforce needs to be developed spanning a broad range of educational levels, job functions, and disciplines, especially in quantum engineering.
- Quantum networking will require a significant manufacturing capability, including new methods of manufacturing to realize its components and subsystems of sufficient quality and performance.
- A key issue for the development of a quantum network is having not only an extremely high probability of success at each step and or/stage of the quantum network but also that the step has high fidelity. The limits of tolerable loss and other errors are much more challenging than for classical optical networks.
- A quantum network must deliver quantum information at rates needed by quantum applications connected to the network. Achieving and sustaining these rates will be challenging.

**Opportunities (Short-Term, 2 - 5 years)**

- The development of an initial common terminology and taxonomy between the various experts required to form the basis of a multidisciplinary quantum networking community. This community should have broad representation from US Government organizations, national laboratories, industry and academia.

**Opportunities (Long-Term 5- 10 years)**

- Quantum networks will enable DOE to harness the full capability of quantum information processing by supporting distributed quantum computing.
- Quantum networks promise to provide long-term security and privacy for the nation's communications and internet connected technologies, such as critical infrastructure.

# 6 Overall Summary and Observations

Quantum networking is a nascent interdisciplinary field in quantum information processing. It is drawing interest from disparate fields such as quantum physics, telecommunications engineering, optical communications, computer science, cyber security, and domain science that have not traditionally worked together. Such collaboration is needed to solve a problem as complex as developing a general-purpose quantum network. However, these communities do not currently have a shared vocabulary or world-view, and many do not understand the DOE's requirements for interconnecting quantum computers and/or quantum sensors. Therefore, this effort will require a relatively long period of collaboration between researchers from these various communities, so that they can achieve a shared understanding of DOE problems and of potential solutions.

Once this shared understanding is achieved, it will be possible to perform a more detailed analysis of alternative concepts to determine whether a single quantum network architecture will satisfy the DOE's needs, or whether multiple (tailored) architectures are required.

# Appendix A: Workshop Attendees

| First Name | Last Name | Affiliation |
|---|---|---|
| Scott | Alexander | Perspecta Labs |
| Yuri | Alexeev | Argonne National Laboratory |
| Fil | Bartoli | ENG/ECCS Division Director, NSF |
| Joe | Britton | Army Research Lab |
| Robert | Broberg | Cisco Systems |
| Michael | Brodsky | US Army Research Lab |
| Benjamin | Brown | DOE SC ASCR |
| Ivan | Burenkov | Joint Quantum Institute at NIST and UMD |
| Stephen | Bush | GE Global Research |
| Mark | Byrd | Southern Illinois University |
| Ryan | Camacho | Brigham Young University |
| Jean-Luc | Cambier | Air Force of Scientific Research |
| Thomas | Chapuran | Perspecta Labs |
| Lali | Chatterjee | DOE HEP |
| Xiaoliang | Chen | UC Davis |
| Tatjana | Curcic | AFOSR |
| Dominique | Dagenais | National Science Foundation |
| Cees | De Laat | University of Amsterdam |
| Jonathan | Dowling | Louisiana State University |
| Yao-Lung (Leo) | Fang | Brookhaven National Lab |
| Eden | Figueroa | Stony Brook University |
| Hal | Finkel | Argonne National Laboratory |
| Warren | Grice | Qubitekk |
| Saikat | Guha | University of Arizona |
| Scott | Hamilton | MIT Lincoln Laboratory |
| James | Harrington | HRL Laboratories |
| Kurt | Jacobs | US Army Research Laboratory |
| Eric | Johnson | National Science Foundation |
| Gregory | Kanter | NuCrypt |
| Rajkumar | Kettimuthu | Argonne National Laboratory |
| Hari | Krovi | Raytheon BBN Technologies |
| Prem | Kumar | Northwestern University |
| Paul | Kwiat | UIUC |
| Nikolai | Lauk | Caltech |
| Randall | Laviolette | ASCR |
| Norbert | Linke | Joint Quantum Institute, University Maryland |
| Alia | Long | Los Alamos National Laboratory |
| Joseph | Lukens | Oak Ridge National Laboratory |
| Joseph | Lykken | Fermi National Accelerator Laboratory |
| Hideo | Mabuchi | Stanford University |
| Sonia | McCarthy | DOE ASCR |
| Grace | Metcalfe | |
| Bogdan | Mihaila | National Science Foundation |
| Alan | Mink | NIST |
| Rich | Mirin | NIST |
| Indermohan | Monga | Lawrence Berkeley National Laboratory |
| Raymond | Newell | Los Alamos National Laboratory |
| Tristan | Nguyen | Air Force of Scientific Research |
| Andrei | Nomerotski | BNL |
| Lucy | Nowell | DOE Office of Science |
| Nicholas | Peters | Oak Ridge National Laboratory |
| Robinson | Pino | DOE Office of Science |
| Bing | Qi | ORNL |
| Gulshan | Rai | Department of Energy, Office of Nuclear Physics |
| Nageswara | Rao | Oak Ridge National Laboratory |
| Akbar | Sayeed | National Science Foundation |
| Thomas | Schenkel | Lawrence Berkeley National Laboratory |
| Kevin | Silverman | NIST-Boulder |
| A. Matthew | Smith | Air Force Research Laboratory |
| Keith | Snail | Army Research Laboratory |
| Daniel | Soh | Sandia National Labs |
| Martin | Suchara | Argonne National Laboratory |
| Ceren | Susut | DOE/ASCR |
| Leandros | Tassiulas | Yale University |
| Donald | Towsley | UMass |
| Brian | Williams | Oak Ridge National Laboratory |
| Alan | Willner | Univ. of Southern California |
| Dantong | Yu | New Jersey Institute of Technology |

# Appendix B: Workshop Agenda

Day 1
| | |
|---|---|
| 07:30 AM – 08:00 AM | Coffee |
| 08:00 AM – 08:05 AM | Logistics: *Julie Webber* |
| 08:05 AM – 08:10 AM | Welcome: *Barb Helland (ASCR Director)* |
| 08:10 AM – 08:45 AM | Workshop Goals & Objectives: *Thomas Ndousse-Fetter* |
| 08:45 AM – 09:00 AM | Workshop Organization: *Prem Kumar, Warren Grice* |
| 09:00 AM – 09:45 AM | Plenary Talk 1: *Chip Elliott* |
| 09:45 AM – 10:15 AM | Plenary Talk 2: *Leandros Tassiulas* |
| 10:15 AM – 10:30 AM | Coffee Break |
| 10:30 AM – 11:00 AM | Quantum Networks: Motivation & Impact Introduction: *Andrei Nomerotski, Nick Peters* |
| 11:00 AM – 12:00 PM | Parallel Breakout Groups |
| 12:00 PM – 01:00 PM | Lunch |
| 12:00 PM – 12:15 PM | Organizing Committee Meeting |
| 01:00 PM – 01:30 PM | Quantum Networks: Network Design Introduction**,** *Tom Chapuran, Don Towsley* |
| 01:30 PM – 02:30 PM | Parallel Breakout Groups Breakout |
| 02:30 PM – 02:45 PM | Coffee Break |
| 02:45 PM – 03:15 PM | Quantum Networks: Devices & Subsystems, *Saikat Guha, Scott Hamilton, Ray Newell.* Introduction: *Edo Waks, Greg Kanter* |
| 03:15 PM – 04:15 PM | Parallel Breakout Groups Breakout |
| 04:15 PM – 04:30 PM | Coffee Break |
| 04:30 PM – 05:00 PM | Quantum Networks Operation & Control Introduction: *Inder Monga, Ben Yoo* |
| 05:00 PM – 06:00 PM | Parallel Breakout Groups Breakout |
| 06:00 PM | Adjourn |
| 07:30 PM – 08:30 PM | Organizing Committee Meeting |

Day 2
| | |
|---|---|
| 07:30 AM – 08:30 AM | Breakfast |
| 08:30 AM – 09:00 AM | Motivation & Impact: *A. Nomerotski, N. Peters* |
| 09:00 AM – 09:30 AM | Network Design: *T. Chapuran, D. Towsley* |
| 09:30 AM – 10:00 AM | Devices & Subsystems: *S. Guha, S. Hamilton, R. Newell* |
| 10:00 AM – 10:30 AM | Operation & Control: *I. Monga, B. Yoo* |
| 10:30 AM – 10:45 AM | Coffee Break |
| 10:45 AM – 11:15 AM | Plenary Talk 3: *Alan Willner* |
| 11:15 AM – 12:15 PM | Cross-cutting Group Discussion: P. *Kumar, W. Grice* |
| 12:15 PM – 12:30 PM | Lunch |
| 12:15 PM – 01:30 PM | Organizing Committee Meeting |
| 01:30 PM – 03:00 PM | Breakout Groups Report: a) Motivation & Impact, b) Network Design, c) Devices and Subsystems, and d) Operations and Control |
| 03:00 PM – 03:15 PM | Coffee Break |

03:15 PM – 04:45 PM          Breakout Groups, Workshop Report
04:45 PM – 05:00 PM          Closing Remarks: *Thomas Ndousse-Fetter*